\documentclass[twocolumn,aps,superscriptaddress,nofootinbib,floatfix,linenumbers]{revtex4}
\usepackage{amsfonts}
\usepackage{float}
\usepackage{placeins}
\usepackage{graphicx}
\usepackage[hidelinks]{hyperref}
\usepackage{color,amsmath,amssymb,bm}
\usepackage{tensor}
\usepackage{epstopdf}

\usepackage[normalem]{ulem}  

\renewcommand{\sout}{\bgroup \color{black} \ULdepth=-.5ex \ULset}

\def\blfootnote{\xdef\@thefnmark{}\@footnotetext}

\newcommand{\beq}{\begin{equation}}
\newcommand{\eeq}{\end{equation}}
\newcommand{\bea}{\begin{eqnarray}}
\newcommand{\eea}{\end{eqnarray}}

\newcommand{\gtsim}{\raisebox{-4pt}{$\,\stackrel{\textstyle >}{\sim}\,$}}

\begin{document}

\title{Assessing the lattice QCD space diffusion coefficient and the thermalization time of charm quark by mean of D meson observables at LHC}

\author{Maria Lucia Sambataro}
\email{sambataro@lns.infn.it}
\affiliation{Department of Physics and Astronomy, University of Catania, Via S. Sofia 64, 1-95125 Catania, Italy}
\affiliation{Laboratori Nazionali del Sud, INFN-LNS, Via S. Sofia 62, I-95123 Catania, Italy}

\author{Vincenzo Minissale}
\affiliation{Department of Physics and Astronomy, University of Catania, Via S. Sofia 64, 1-95125 Catania, Italy}
\affiliation{INFN Sezione di Catania, Via S. Sofia 64, 95123 Catania, Italy}

\author{Salvatore Plumari}
\affiliation{Department of Physics and Astronomy, University of Catania, Via S. Sofia 64, 1-95125 Catania, Italy}
\affiliation{Laboratori Nazionali del Sud, INFN-LNS, Via S. Sofia 62, I-95123 Catania, Italy}

\author{Vincenzo Greco}
\affiliation{Department of Physics and Astronomy, University of Catania, Via S. Sofia 64, 1-95125 Catania, Italy}
\affiliation{Laboratori Nazionali del Sud, INFN-LNS, Via S. Sofia 62, I-95123 Catania, Italy}

\date{\today}

\begin{abstract}

A central goal in the study of heavy-flavour production is to determine the interaction strength between Heavy Quarks (HQs) and the Quark-Gluon Plasma (QGP), quantified by the spatial diffusion coefficient $D_s(T)$. Recent lattice QCD (lQCD) results with dynamical fermions suggest a remarkably low value of $2\pi T D_s \approx 1$ at $T=T_c$ for charm quarks - significantly lower than both quenched QCD estimates and most phenomenological models - which typically yield $2\pi T D_s \approx 3.5 - 5$. This discrepancy raises the question of whether such a small $D_s(T)$, corresponding to a thermalization time $\tau_{th} \approx 1 - 1.5$ fm/c, is compatible with experimental measurements of key observables like the nuclear modification factor  $R_{AA}$, the elliptic and triangular flow coefficients $v_2$ and $v_3$ for D mesons. Using an event-by-event Langevin transport framework, we analyze several scenarios and highlight the pivotal role played by the momentum dependence of the drag coefficient $A(p) = \tau_{th}^{-1}(p)$. Our findings show that a small $2\pi T D_s (p\rightarrow 0)\approx 1 - 2$ values can align with experimental data \emph{only} if a significant momentum dependence in $\tau_{th}(p)=1/A(p)$ is included, as predicted by T-matrix approaches, or by the extended Quasi-Particle Model (QPMp). In contrast, assuming a momentum-independent $\tau_{th} = M_c D_s^{\text{lQCD}} / T$, it fails to reproduce the observed phenomenology. Furthermore, a short thermalization time of $\tau_{th} \approx 1.5$ fm/c implies a loss of sensitivity of the final-state observables to the initial charm-quark momentum distribution up $p_T \approx M_c$, suggesting a possible universal behavior driven by a dynamical attractor.

\end{abstract}

\maketitle

\section{Introduction}

Charm and bottom quarks, collectively referred to as Heavy Quarks (HQs), serve as powerful probes of the QCD medium created in Ultrarelativistic Heavy-Ion Collisions (URHICs) \cite{Dong:2019unq,Rapp:2008qc,He:2022ywp}. These particles are created by initial hard processes and, during the fireball evolution, the number of charm and bottom quarks is almost conserved. This is mainly due to their large mass with respect to the typical temperature reached at RHIC and LHC, implying a negligible thermal production. HQs will likely be relevant probes also at the FAIR/SPS energy  thanks to the CBM and NA60+ experiments\cite{Inghirami:2018vqd,NA60:2022sze}.The propagation of HQs in the Quark-Gluon Plasma (QGP) has frequently been modeled in a Brownian motion since the large mass of HQs should generally lead to collisions with small momentum transferred. Under these conditions, the Boltzmann transport equation simplifies to Fokker-Planck dynamics, providing a significant reduction in the complexity of in-medium dynamics. This framework has been extensively used to calculate key observables \cite{Dong:2019unq,He:2022ywp,Scardina:2017ipo,vanHees:2005wb,vanHees:2007me,Gossiaux:2008jv,Das:2009vy,Alberico:2011zy,Uphoff:2012gb,Lang:2012nqy,Song:2015sfa,Song:2015ykw,Das:2013kea,Cao:2015hia,Das:2015ana,Cao:2017hhk,Das:2017dsh,Jamal:2020fxo,Ruggieri:2018rzi,Sun:2019fud,Cao:2018ews,Rapp:2018qla,Sambataro:2023tlv,Sambataro:2022sns,Plumari:2019hzp}, such as the nuclear suppression factor $(R_{AA})$ which describes how the HQs spectra are modified in AA collisions w.r.t. the pp collisions and also the elliptic flow $v_2=\langle cos(2\phi_p) \rangle$, a measure of the anisotropy in the angular distribution of heavy flavour hadrons. 
Recently, new calculations of the spatial diffusion coefficient $D_s$ obtained for finite HQs mass supplementing lQCD which include dynamical fermions with Non-Relativistic Effective Fields Theory (NREFT), have paved the way to realistic estimates of $D_s$ and HQs thermalization relaxation time for full QCD. In particular, the new lQCD data are significant smaller than the previous ones in quenched approximation and exhibit a weak mass dependence going from charm to infinite mass limit \cite{Altenkort:2023oms,Altenkort:2023eav,HotQCD:2025fbd}. The small $2 \pi T D_s \approx 1$ of lQCD/NREFT suggests a thermalization time $\tau_{th}\approx 1.4 \, fm/c$ at $T \approx 1.2$ $T_c$ at charm mass scale compatible for $p\rightarrow 0$ to the AdS/CFT predictions \cite{Gubser:2006qh,Horowitz:2015dta,Casalderrey-Solana:2006fio} and significantly smaller than several phenomenological estimates \cite{Cao:2018ews}. A main motivation of this work is to investigate whether such low values of $D_s$ are compatible with a description of $R_{AA}$ and $v_n$ in agreement with the available experimental data. 
Several effective models have been developed to incorporate non-perturbative aspects of the interaction between HQs and plasma particles. There are approaches in which pQCD predictions are improved by incorporating Hard Thermal Loop (HTL) resummations \cite{Gossiaux:2008jv,Nahrgang:2014vza,Nahrgang:2013saa,Alberico:2011zy}, or approaches developed in the past decade like Quasi-Particle Models (QPMs) \cite{Peshier:2002ww, Plumari:2011mk, Song:2015ykw, Liu:2023rfi, Soloveva:2023tvj, Peshier:2005pp, Berrehrah:2015vhe, Berrehrah:2016vzw, Grishmanovskii:2025mnc, Sambataro:2024mkr}. The QPMs introduce effective temperature-dependent masses and have been able to reproduce the thermodynamic properties obtained in lQCD and they have provided a more satisfactory description of the HQ spatial diffusion coefficient $D_s$, with estimations in agreement with lQCD results in quenched approximation \cite{Banerjee:2011ra, Kaczmarek:2014jga, Francis:2015daa, Brambilla:2020siz}. An alternative framework is based on the TAMU formalism (TAMU), which models the HQ-medium interactions through a potential kernel fitted to the heavy-quark free energy obtained from lattice QCD and more recently to the Wilson correlator, leading to a stronger momentum dependence of the drag coefficient \cite{Tang:2023lcn,Liu:2017qah,Tang:2023tkm}. Recently, an extended version of the $QPM$, named $QPM_p$, has been developed to include momentum dependent masses that recover the current quark masses in the limit of infinite momentum.  
This extended approach is able to correctly reproduce not only the lQCD Equation of State (EoS) but also both light and strange quark susceptibilities which are underestimated in the standard QPM with temperature-only dependent masses \cite{Sambataro:2024mkr}. Within the $QPM_p$ a $D_s(T)$ with a strong non-perturbative behavior is predicted in the small T region with a sizeable mass dependence going from charm to infinite mass limit than the one exhibited in lQCD data. In particular, the $D_s(T)$ in $QPM_p$ for bottom quark is close to the recent lQCD/NREFT data near $T_c$ ($2\pi T D_s \approx 1$) but the charm quark $D_s(T)$ in $QPM_p$ significantly deviates already at $T \gtsim  \, T_c$ with $2\pi T D_s \approx 2.5 $, suggesting a larger thermalization time $\tau_{th} \approx 2.5 \, fm/c$ wrt the new lQCD estimates. In order to explore whether the small values of new lQCD $D_s$ are compatible to the experimental data of $R_{AA}$ and $v_n$ and the role which the momentum dependence plays in this context, we explore different scenarios by comparing the D-meson $R_{AA}$ and $v_{2,3}$ obtained with the interaction modeled via the $QPM_p$ to the results obtained with the interaction scaled to new lQCD data and assuming a constant drag $A=1/(M D_s)$, or a p-dependence as in the T-Matrix. We have developed a Langevin transport approach which include event-by-event fluctuations in the initial state. In this approach, the hadronization process is modeled by combining standard coalescence and fragmentation that has provided successful predictions of the large baryon-to-meson ratios at intermediate $p_T$ and the constituent-quark number scaling of the elliptic flow $v_2(p_T)$ for light hadrons production  \cite{Fries:2008hs,Greco:2003mm,Greco:2003vf,Minissale:2015zwa,Fries:2025jfi}. In recent years, coalescence models have been also applied to the study of heavy-flavor (HF) hadron chemistry in $AA$ at RHIC and LHC energies, providing predictions also for multi-charm hadrons in different collisions systems \cite{Greco:2003vf, Plumari:2017ntm, Minissale:2023dct}. Furthermore, it has been recently used to describe the hadronization mechanism also in small system like $pp/pA$ collisions and in light-ion collisions as for the upcoming experiments expected to be performed at LHC \cite{Minissale:2020bif,Minissale:2024gxx}. The paper is organized as follows: in the section II, we describe the event-by-event initial conditions of partons implemented in our approach. In section III, we describe the event-by-event relativistic transport Boltzmann approach to describe the bulk evolution coupled to an event-by-event Langevin approach used to study the charm quark dynamics in the QGP. In section IV, we presents the results on $R_{AA}$ and $v_{2,3}$ of D mesons, highlighting the key role of the momentum dependence of the interaction in describing the experimental observables. Finally, in section VI we conclude with a summary and final remarks.

\section{Event by event transport approach for charm dynamics}

In order to describe the initial conditions of partons in coordinate space, we have employed a MonteCarlo Glauber Model modified according the wounded quark model to account for initial state fluctuations as already used within the Boltzmann and hydrodynamics framework in \cite{Sun:2019gxg, Sambataro:2022sns, Sambataro:2022xzx, Petersen:2010cw, Holopainen:2010gz, Schenke:2011bn, Gale:2012rq}. This event-by-event transport approach give us the possibility to study not only the $v_2$ but also higher order anisotropic flows, such as the triangular flow $v_3$. In this approach, the nucleons in $Pb$ nuclei are distributed according to the standard Woods–Saxon distribution while the three constituent quarks of each nucleon are randomly distributed within each nucleon according to the distribution $dN/dr=\frac{r^2}{r_0^3}e^{-r/r_0}$ with $r_0\!=\!0.3$ $fm$. The center of mass of the three quarks is then translated to the position of the nucleon and the probability that each quark pair from target and projectile can collide or not is described by $p = e^{-\pi r^2/\sigma_{qq}}$ with $\sigma_{qq}\! =\! 13.6 \mbox{ mb}$ in 5.02 ATeV $Pb + Pb$ collision \cite{Bozek:2016kpf}. At the end, the total initial parton distribution can be expressed in the following form:

\begin{equation}\label{profile_qu}
    \frac{dN}{d^2\textbf{x}_\perp d\eta}=\sum_{i=1}^{N_{part}}n_i\rho_\perp(\textbf{x}_\perp-\textbf{x}_i)\rho(\eta)
\end{equation}

where $n_i$ is the number of partons produced by each partecipant quark. Furthermore, $N_{part}$ and $\textbf{x}_i$ represent the total number of participant quarks and their transverse position respectively. In particular, the profile for the spatial rapidity distribution $\rho(\eta)$ is given by longitudinal boost invariance in pseudo-rapidity while the transverse position profile $\rho_\perp(\textbf{x}_\perp-\textbf{x}_i)$ is expressed by:

\begin{equation}\label{profile_gau}
   \rho_\perp(\textbf{x}_\perp-\textbf{x}_i)=\frac{1}{2\pi\sigma^2}e^{-\frac{(\textbf{x}_\perp-\textbf{x}_i)^2}{2\sigma^2}}.
\end{equation}

In Eq. \ref{profile_qu}, the number of partons generated in each event can be expressed by $n_i \!=\! n_0 N$ with a fixed $n_0$ to reproduce the same final charged particle multiplicity measured in experiments. Since the number of particles produced in $pp$ collisions fluctuates according to a negative binomial distribution, $N$ is sampled by the following expression:

\begin{equation}
  P(N)=\frac{\Gamma(N+k)\bar n^N k^k}{\Gamma(k) N!(\bar n+k)^{N+k}}  
\end{equation}

where we have fixed $k= 0.224$, $\bar{n} = 1.621$ and $n_0 \approx 1.88$ to correctly reproduce the charged particles distribution by ALICE Collaboration. 
In our model, regarding the particle distribution in the momentum space, we distribute \textit{soft partons}, i.e. quarks and gluons with an initial thermal equilibrium distribution, according to the following expression:
\begin{eqnarray}
\label{distr_light}
    & &\frac{dN}{d^2\mathbf{x}_\perp d \eta}=\frac{g \tau_0}{(2\pi)^2} p_T m_T \times \\ \nonumber
    &\times &\cosh{(\eta-y)} \exp\left(-\frac{m_T\cosh{(\eta - y)}}{T(\mathbf{x}_\perp, \eta)}\right) dp_T dy
\end{eqnarray}

where $\tau_0=0.3 fm/c$ is the initial time of plasma particles evolution taken from standard hydro approach,  $g=2\times8+3\times2\times6=52$ is the degree of freedom of partons (three flavor quarks and gluons) and $m_T=\sqrt{m^2+p_T^2}$ is the light partons transverse mass. On the other hand, we consider a \textit{hard} component of minijets consisting in the products of initial binary pQCD collision whose transverse momentum distributions at mid-rapidity are taken from spectra of CUJET Collaboration \cite{Xu:2015bbz} for $pp$ collisions at 5.02 TeV. 
We initialize the charm quark distribution in coordinate space according to the binary collision profile $N_{coll}$ from the Monte Carlo Glauber model. In the momentum space, the distribution of HQs is fixed to the D-meson spectra in $pp$ collisions after fragmentation which is given by Fixed Order + Next-to-Leading Log (FONLL) calculations ~\cite{Cacciari:2012ny}. For more details, see Ref.~\cite{Scardina:2017ipo}. Finally, in our model the charm quark can hadronize via an hybrid approach of hadronization by coalescence
plus fragmentation.  The key ingredients of the hadronization approach are presented in details in several papers \cite{Plumari:2017ntm, Minissale:2023dct, Minissale:2015zwa,Cao:2015hia,Gossiaux:2009mk,Oh:2009zj,Minissale:2024gxx}.

\section{Transport equations for charm and bulk}

The results shown in this paper have been obtained employing an event-by-event relativistic transport approach. In particular, the evolution of QGP bulk matter is given by the relativistic Boltzmann equation at fixed $\eta/s$ as described in details in our previous works in which we studied the dynamics of heavy-ion collisions at both RHIC and LHC energies \cite{Plumari:2012ep, Ruggieri:2013bda, Scardina:2012mik, Ruggieri:2013ova, Puglisi:2014sha,  Scardina:2014gxa, Plumari:2015sia, Plumari:2015cfa, Scardina:2017ipo, Plumari:2019gwq, Sun:2019gxg, Sambataro:2020pge}. Therefore, the evolution of the gluon and light quark is described by the following equations: 

\begin{equation}
p^{\mu}_{j} \partial_{\mu}f_{j}(x,p)= {\cal C}[f_{j}](x_{j},p_{j}) \, \, for \, \, j=g,q \label{eq:Boltzmann}
\end{equation}

where ${\cal{C}}[f_j](x,p)$ is the relativistic Boltzmann-like collision integral which accounts for $2\rightarrow2$ scattering processes. In our approach, we simulate a medium with a fixed $\eta/s = 0.1$ by calibrating the collision integral $C[f_{g,q}]$ to viscous hydrodynamics. For more details about the relativistic transport approach at fixed $\eta/s$, see Refs~\cite{Plumari:2019gwq, Plumari:2015cfa, Ruggieri:2013ova, Plumari:2012xz}. In this paper, we treat the HQs propagation in the QGP in the brownian motion approximation by means of the Fokker-Planck equation which is equivalent to a set of stochastic differential equation, i.e. Langevin equations.

The relativistic equations of HQs motion can be expressed by the following equations:

\begin{eqnarray}
& & dx_j= \frac{p_j}{E}dt \label{eq:x_lang} \\
& & dp_j= -\Gamma p_j dt + \sqrt{dt} C_{j,k} \rho_k \label{eq:y_lang}
\end{eqnarray}

where $\Gamma$ and $C_{j,k}$ are directly connected to the \textit{drag} and \textit{diffusion} coefficients respectively. In particular, the first term correspond to a “deterministic” part describing the momentum variation due to interactions with the bulk medium while the second term is the “stochastic” part describing the fluctuations in terms of independent Gaussian-normal distributed random variables $\rho_k$ \cite{Rapp:2008qc}. In order to define the stochastic integral, we need to specify the momentum argument in the covariance matrix $C_{j,k}(t,x,p+\xi dp)$ as discussed in Ref. \cite{Rapp:2008qc}. In the following, we choose the post-Ito realization $\xi=1$.
\begin{figure}[ht!]\centering
\includegraphics[width=1.0\linewidth]{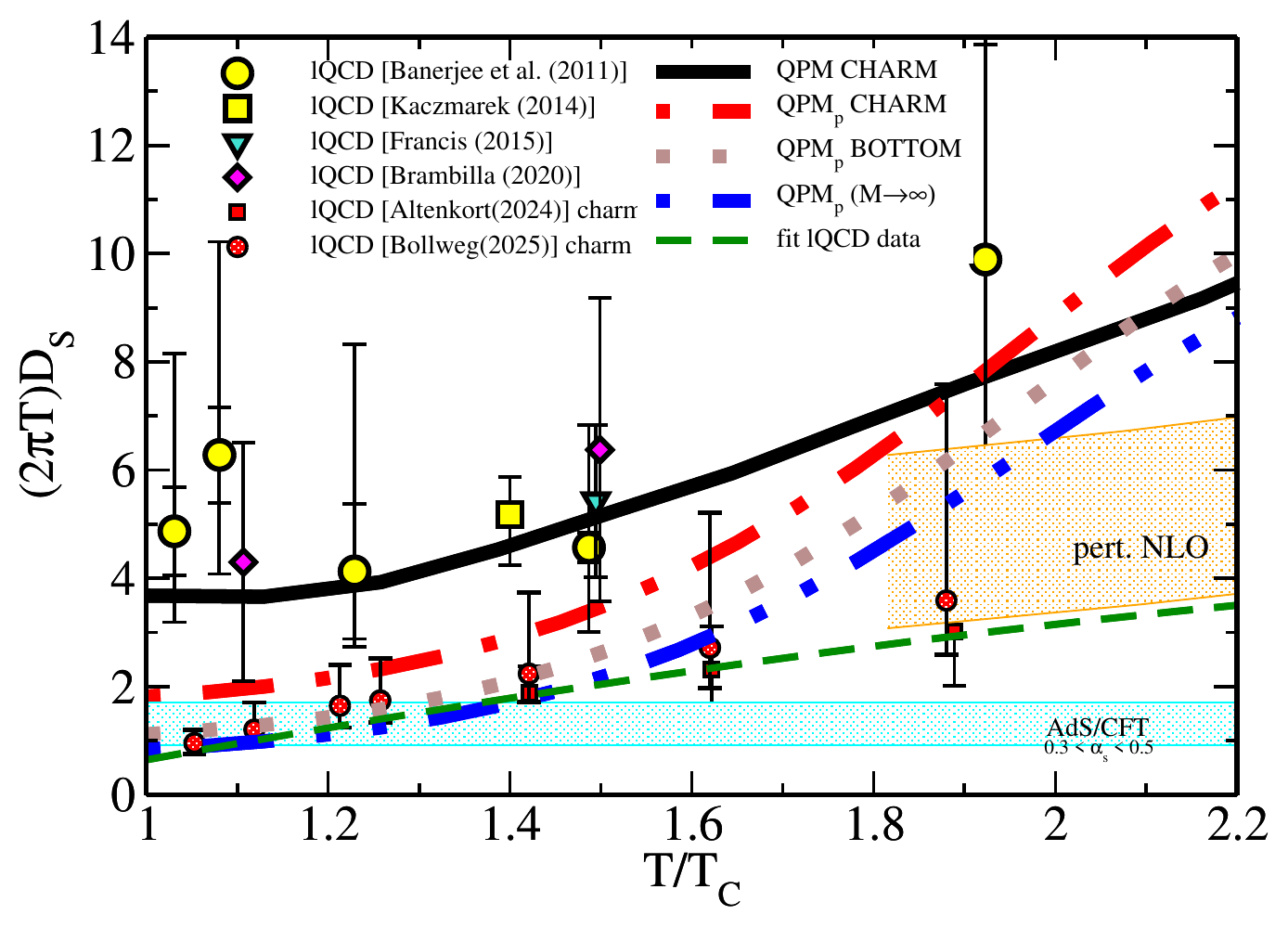}
\caption{Spatial diffusion coefficient $2\pi T D_s$ in $QPM_p$ and $T$-matrix approaches for charm quark compared to available lQCD data.}\label{fig-Ds}
\end{figure}
Therefore, we can write $C_{j,k}\!=\!\sqrt{2D_p(E)}\delta_{jk}$, where $D_p$ is the diffusion coefficient in momentum space, and the fluctuation-dissipation relation takes the simple form $D_p(p)=A(p)E(p)T$. Note that the drag $A$ and diffusion $D_p$ coefficients are functions of $T(x)$ which is the bulk temperature extracted by the Boltzmann equation via a coarse-graining procedure.
In the static limit $p\rightarrow 0$, we remind that we can write $A(p\rightarrow 0)=\gamma$ and the diffusion coefficient $D_p$ in momentum space can be related to a \textit{spatial diffusion coefficient $D_s$} via

\begin{eqnarray}\label{eq:Ds}
    D_s=\frac{T^2}{D_p}=\frac{T}{M_{HQ}\gamma}=\frac{T}{M_{HQ}}\tau_{th}.
\end{eqnarray}

We also consider another approach recently developed, $QPM_p$, where the drag coefficient at finite momenta $A(p,T)$ is evaluated from the scattering matrices $|\cal{M_{HQ}}|$ computed at tree level but with effective vertex coupling $g(T)$ obtained from a fit to lQCD thermodynamics of the $QPM_p$\cite{Borsanyi:2016ksw}. The resulting coupling is significantly larger than the one in pQCD approach, particularly as $T \rightarrow T_c$ (see details in Refs~\cite{Sambataro:2024mkr}). The $QPM_p$, incorporating p-dependent parton masses as entailed by the QCD asymptotic freedom, has been shown to lead to a good description of lQCD Equation of State, but also of the light, strange and charm quark susceptibilities. This approach implies a $D_s(p\rightarrow 0)$ for charm quark close to the new lQCD/NREFT data near $T_c$ (but still at least a $50\%$ larger) and  for $M_{HQ} \rightarrow \infty $ and bottom quark a $D_s$ in very close agreement to lQCD data; see in fig. \ref{fig-Ds} red dash-dotted line (charm), marron dashed line (bottom) and the double-dot dashed blue line ($M_{HQ} \rightarrow \infty $). In fig. \ref{fig-Ds}, we show the $2\pi T D_s$ in $QPM_p$ for charm quark (red dot-dashed line), bottom quark (dotted grey line) and  $M_{HQ} \rightarrow \infty$ (double dot-dashed blue line) in comparison to the lQCD data in quenched approximation and the latest ones including dynamical fermions (red circles and squares). We also show the higher values of $2\pi T D_s$ obtained in standard $QPM$ (black solid line) with temperature-only dependent mass. The small value of lQCD $D_s$ which slightly increases with the temperature, implies a thermalization time $\tau_{th}$ for charm quark compatible to the AdS/CFT estimates (cyan band) \cite{Gubser:2006qh,Horowitz:2015dta,Casalderrey-Solana:2006fio} especially near $T_c$. In the same figure, we also include the perturbative Next-to-Leading Order predictions for $2 \pi T D_s$ \cite{Caron-Huot:2007rwy}.

\section{Results}

The recent results of $D_s$ provided by lQCD calculation which include dynamical fermions are significantly smaller than the ones evaluated in the past years for a quenched medium, leading to a thermalization time for the charm quark which at $T_c$ decreases in the $p\rightarrow 0$ limit from a value of $\tau_{th} \approx 4-5 fm/c$ to a value of $\tau_{th} \approx 1.2 fm/c$. This small value also implies that the charm quark could reach the thermal equilibrium with the hot medium and could even be described within a  viscous hydro approach.
\begin{figure}[ht]\centering
\includegraphics[width=0.8\linewidth]{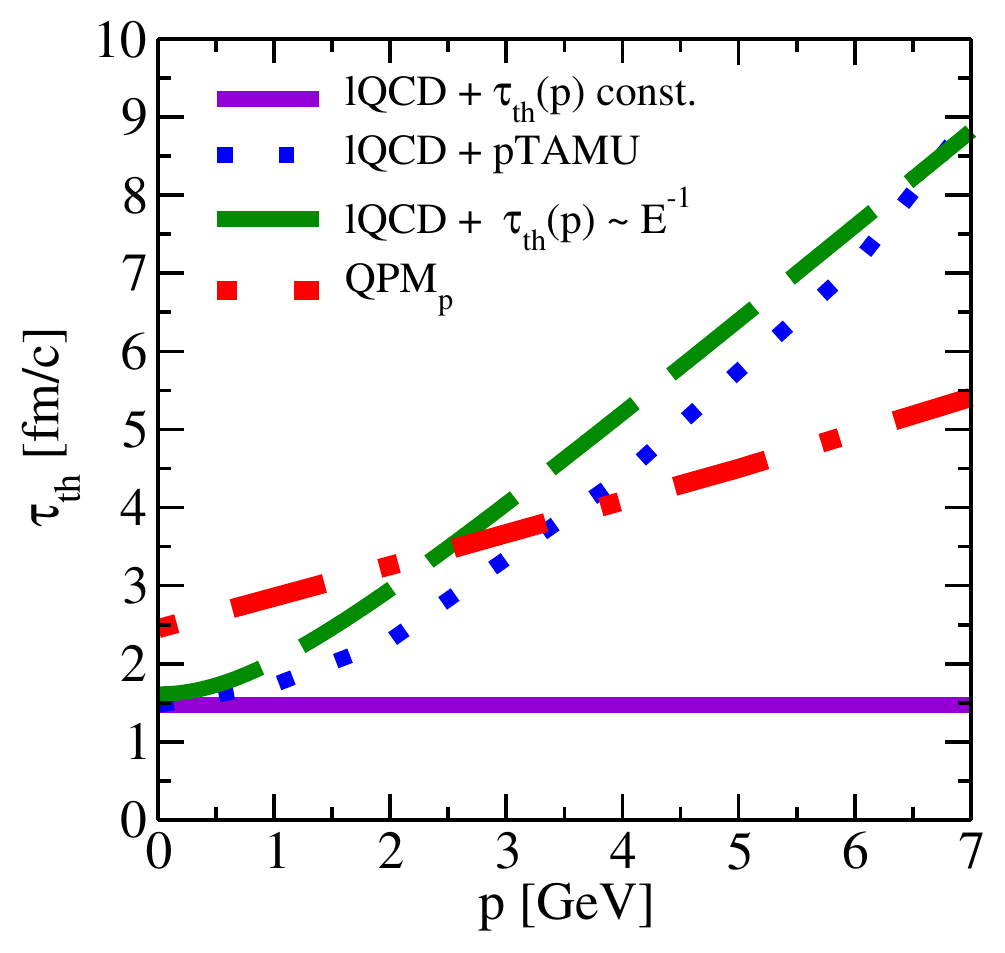}
\caption{Thermalization time $\tau_{th}$ as function of charm momentum $p$ in the \textit{Case 1} of interaction scaled to lQCD $D_s(T)(p=0)$ and drag from FDT relation $A=T/MD_s$ (violet solid line), in the \textit{Case 2} of $D_s(T)$ as in lQCD  with TAMU momentum dependence (blue dotted line), in the \textit{Case 3} with drag from FDT at finite momenta $A=T/ED_s$ (green dashed line) and in the \textit{Case 4} for full $QPM_p$ (red dot-dashed line). }\label{fig-tau}
\end{figure}
\begin{figure}[ht]\centering
\includegraphics[width=1.0\linewidth]{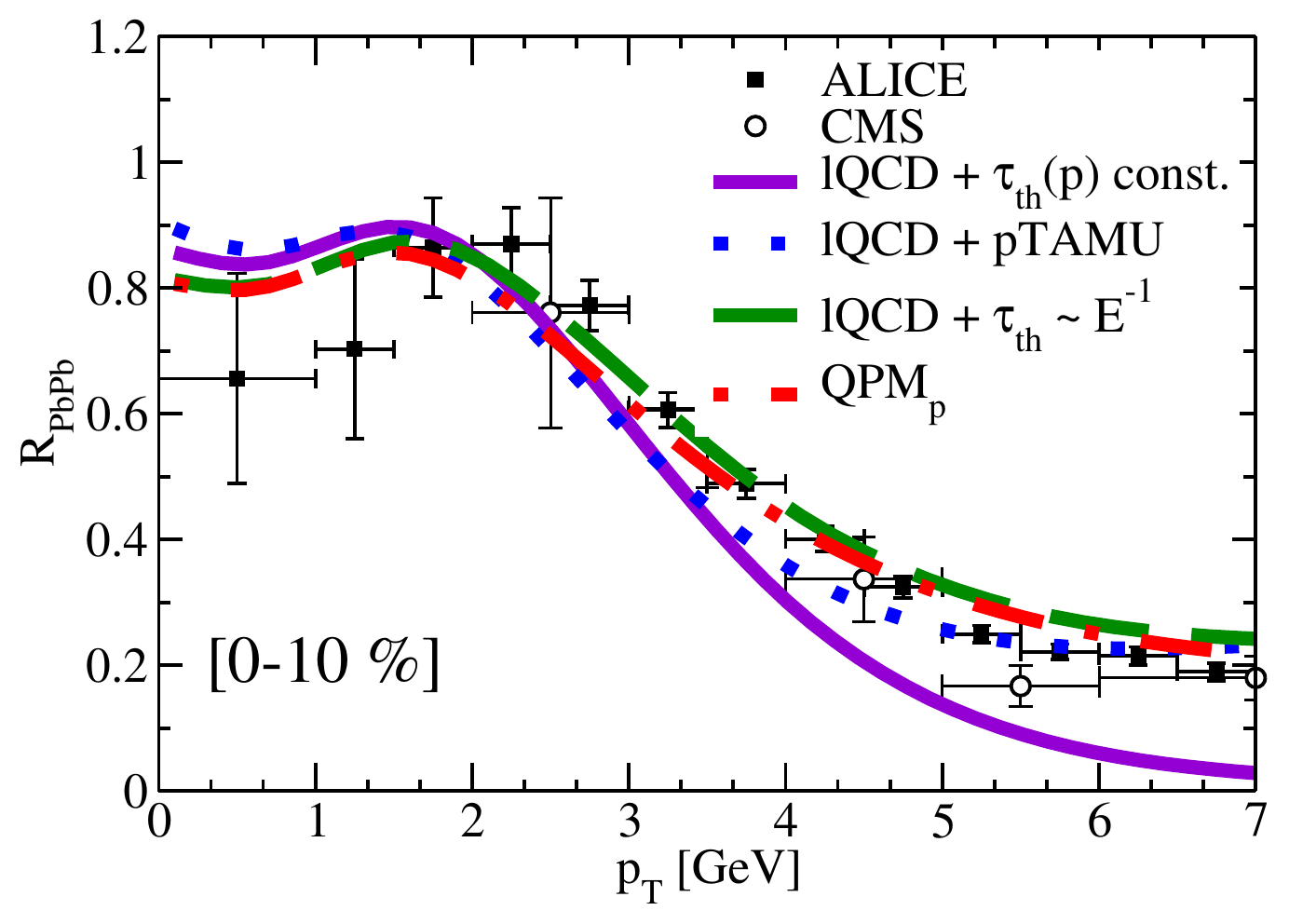}
\caption{$R_{AA}(p_T)$ of D mesons a $0-10 \%$ for the different cases studied. Experimental data from ALICE Coll.\cite{ALICE:2021rxa} and CMS Coll. \cite{CMS:2017qjw}. Same legend of Fig. \ref{fig-tau}.}\label{fig-Raa010}
\end{figure}
\begin{figure}[ht]\centering
\includegraphics[width=1.0\linewidth]{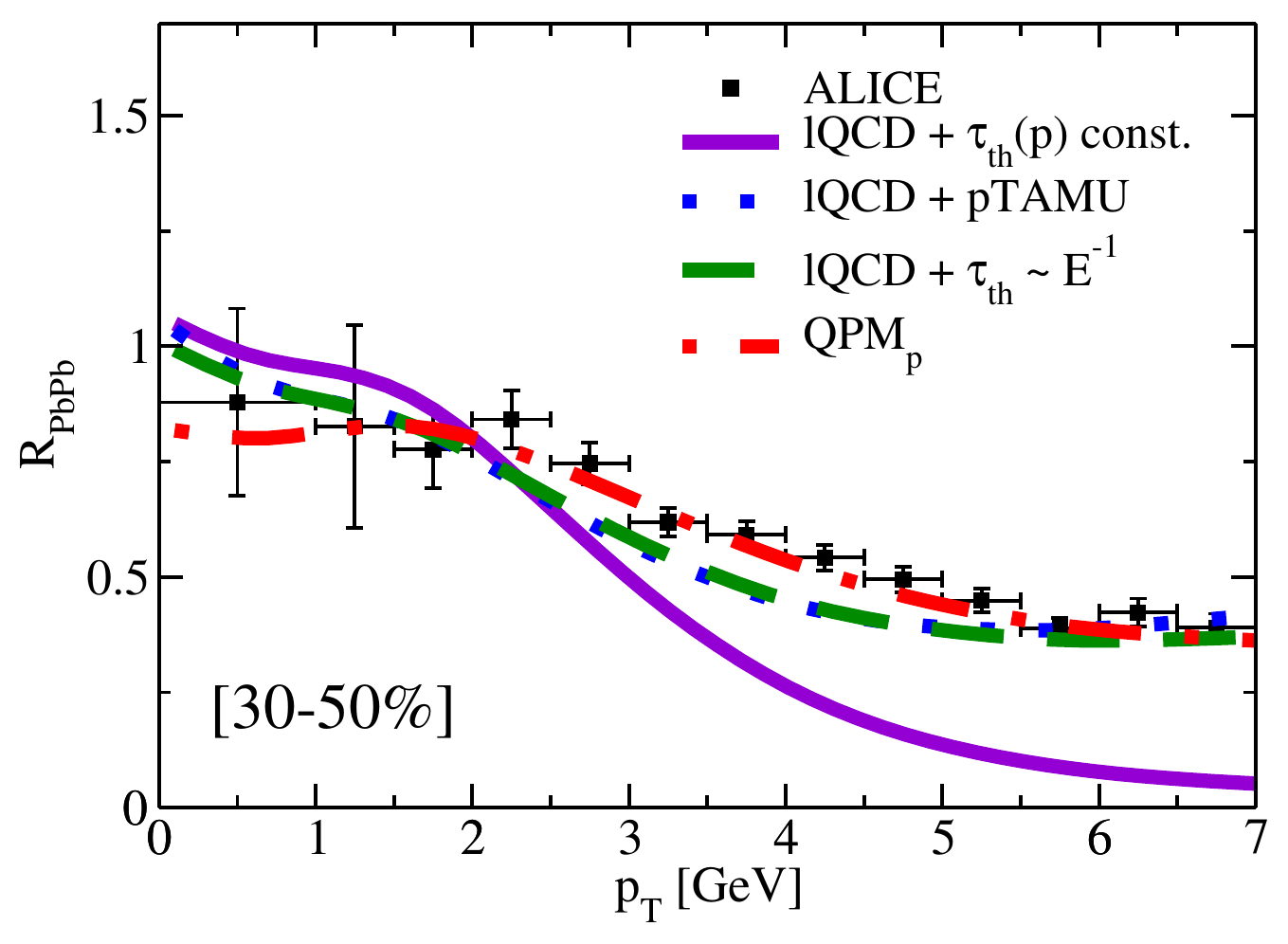}
\caption{$R_{AA}(p_T)$ of D mesons at $30-50 \, \%$ for the different cases studied. Experimental data from ALICE Coll.\cite{ALICE:2021rxa}. Same legend of Fig. \ref{fig-tau}.}\label{fig-Raa3050}
\end{figure}
In this section we analyze if such a small value of $D_s$ even for charm quark can be compatible to experimental data of $R_{AA}$ and $v_{2,3}$. In particular, we well emphatize the key relevance of momentum dependence of $D_s$ and/or drag coefficient $A$ implied by phenomenological models like $QPM_p$ or TAMU or even simply by the Fluctuation Dissipation Theorem (FDT). Hence, we have considered charm quark with mass $M_c = 1.4 \, GeV$ and investigated four different scenarios:

\begin{itemize}
    \item \textit{Case 1} (lQCD + $\tau_{th}(p)$ const.): The $D_s$ as in lQCD data \cite{Altenkort:2023eav,HotQCD:2025fbd} and the drag is obtained from FDT with $A=\frac{T}{M D_s}$. In other words, this is the case where drag and diffusion are those that can be inferred from lQCD and no dependence on momentum is assumed implying a constant thermalization time.
    
    \item \textit{Case 2} (lQCD + pTAMU): The $D_s$ as in the new lQCD data, i.e. the drag coefficient $A(T)$ is extracted from the $D_s$ in lQCD (green dashed line in Fig. \ref{fig-Ds}) by Eq. \ref{eq:Ds}, but imposing a p-dependence of drag $A(p)$ (hence of $D_s$, see Eq. \ref{eq:Ds}) as in the TAMU approach from Ref. \cite{Tang:2023tkm}.

    \item \textit{Case 3} (lQCD + $\tau_{th}\sim E^{-1}$): The $D_s$ as in lQCD data but without momentum dependence while the drag is obtained from FDT extended to finite momentum with $A(p)=\frac{T}{E D_s}$. We note that such a case implies a $D_s(p)$ nearly momentum independent.
    
    \item \textit{Case 4} ($QPM_p$): The interaction of HQs with the plasma as function of momentum is taken from the $QPM_p$ approach \cite{Sambataro:2024mkr}, i.e. the drag coefficient $A(p,T)$ is evaluated from the scattering matrices $|M_{HQ}|^2$ with the effective coupling $g(T)$ of the extended $QPM_p$.
    
\end{itemize}

\begin{figure}[ht]\centering
\includegraphics[width=1.0\linewidth]{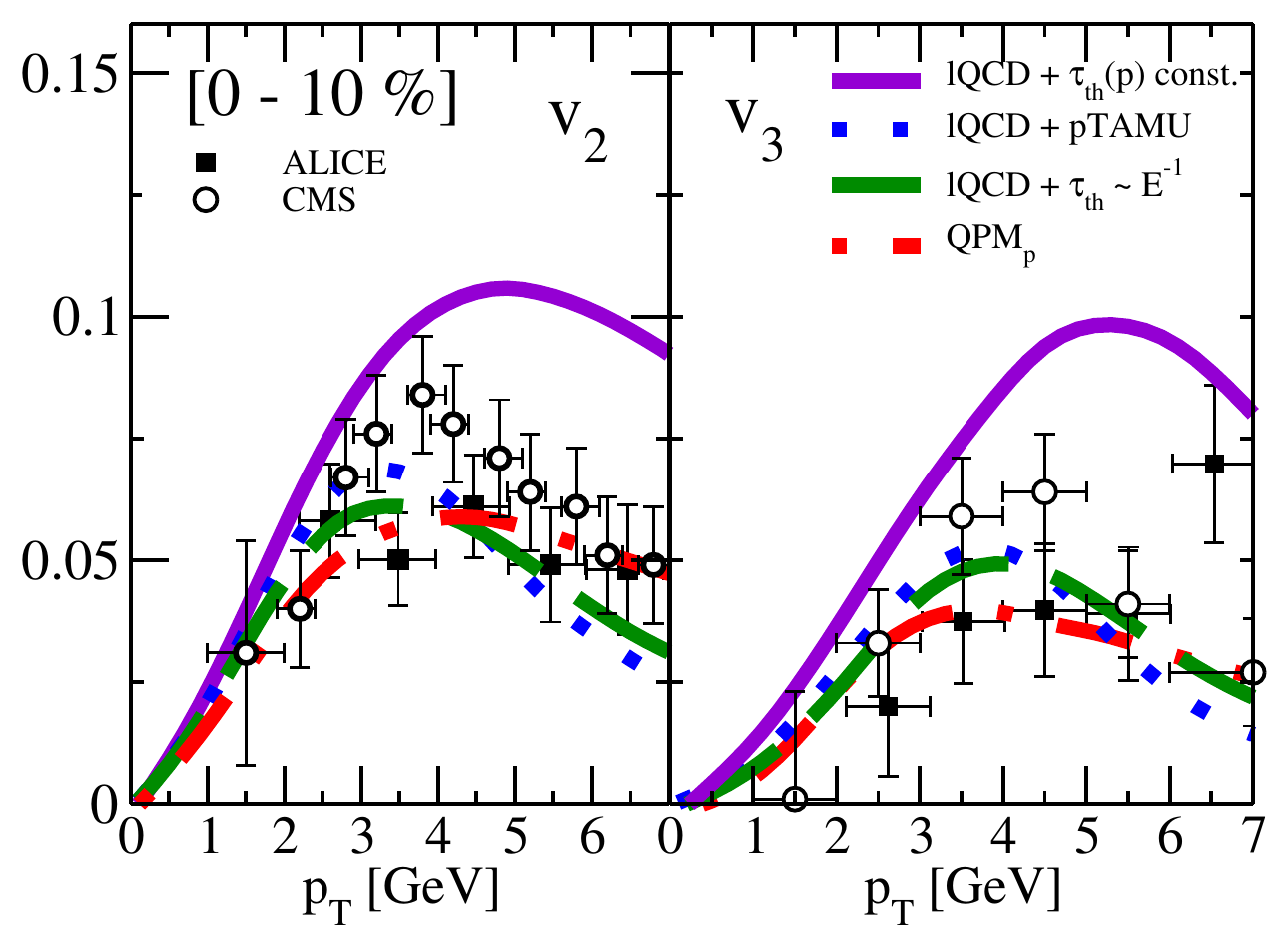}
\caption{Elliptic flow $v_{2}(p_T)$ and triangular flow $v_{3}(p_T)$ of D mesons a $0-10 \%$ for the different cases studied. Experimental data are from ALICE Coll.\cite{ALICE:2020iug} (black squares) and CMS Coll. \cite{CMS:2020bnz} (white circles). Same legend of Fig. \ref{fig-tau}.}\label{fig-v2010}
\end{figure}

\begin{figure}[ht]\centering
\includegraphics[width=1.0\linewidth]{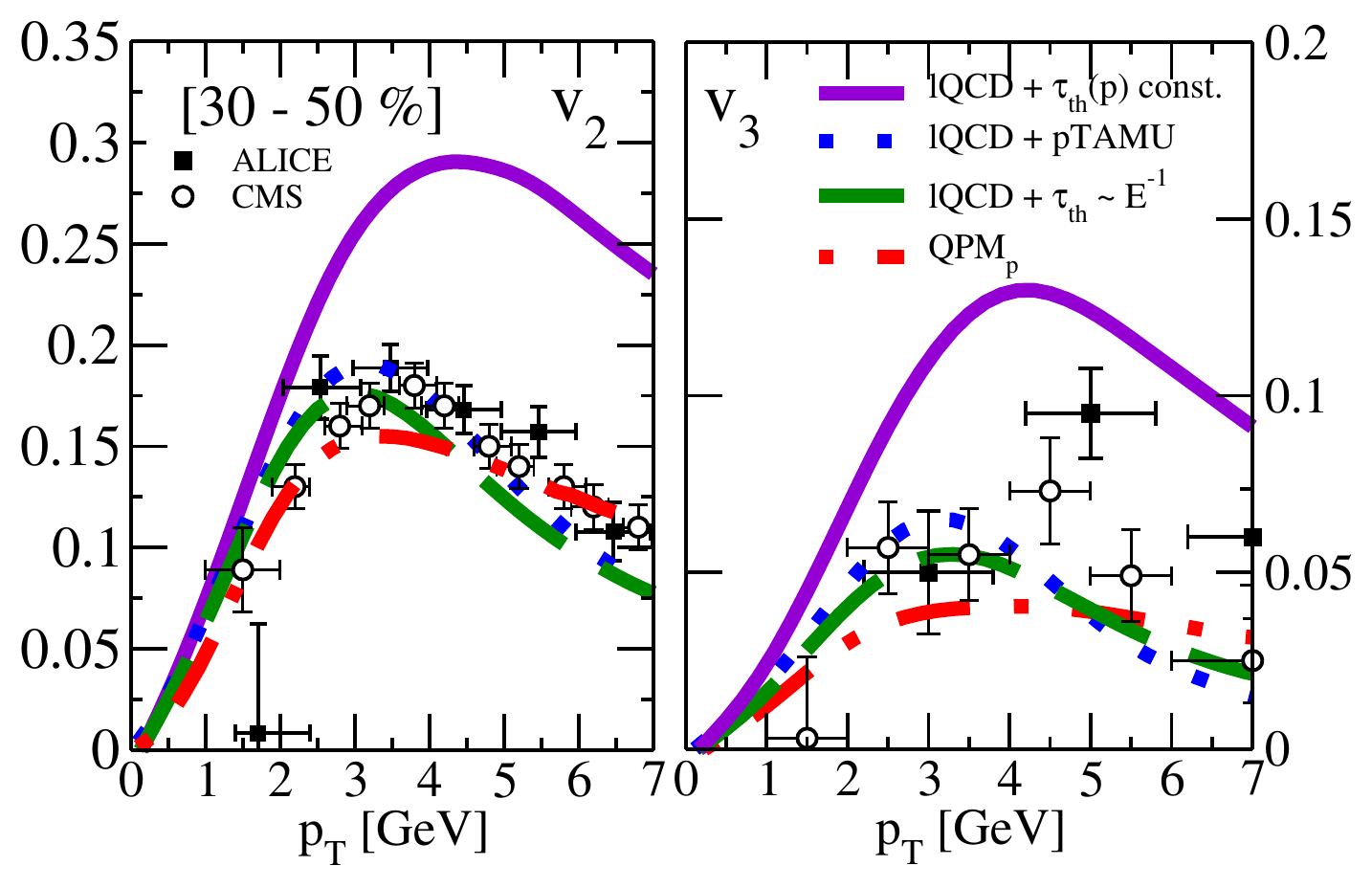}
\caption{Elliptic flow $v_{2}(p_T)$ and triangular flow $v_{3}(p_T)$ of D mesons a $30-50 \%$ for the different cases studied. Experimental data are from ALICE Coll.\cite{ALICE:2020iug} (black squares) and CMS Coll. \cite{CMS:2020bnz} (white circles). Same legend of Fig. \ref{fig-tau}.}\label{fig-v23050}
\end{figure}

\begin{figure}[ht]\centering
\includegraphics[width=1.0\linewidth]{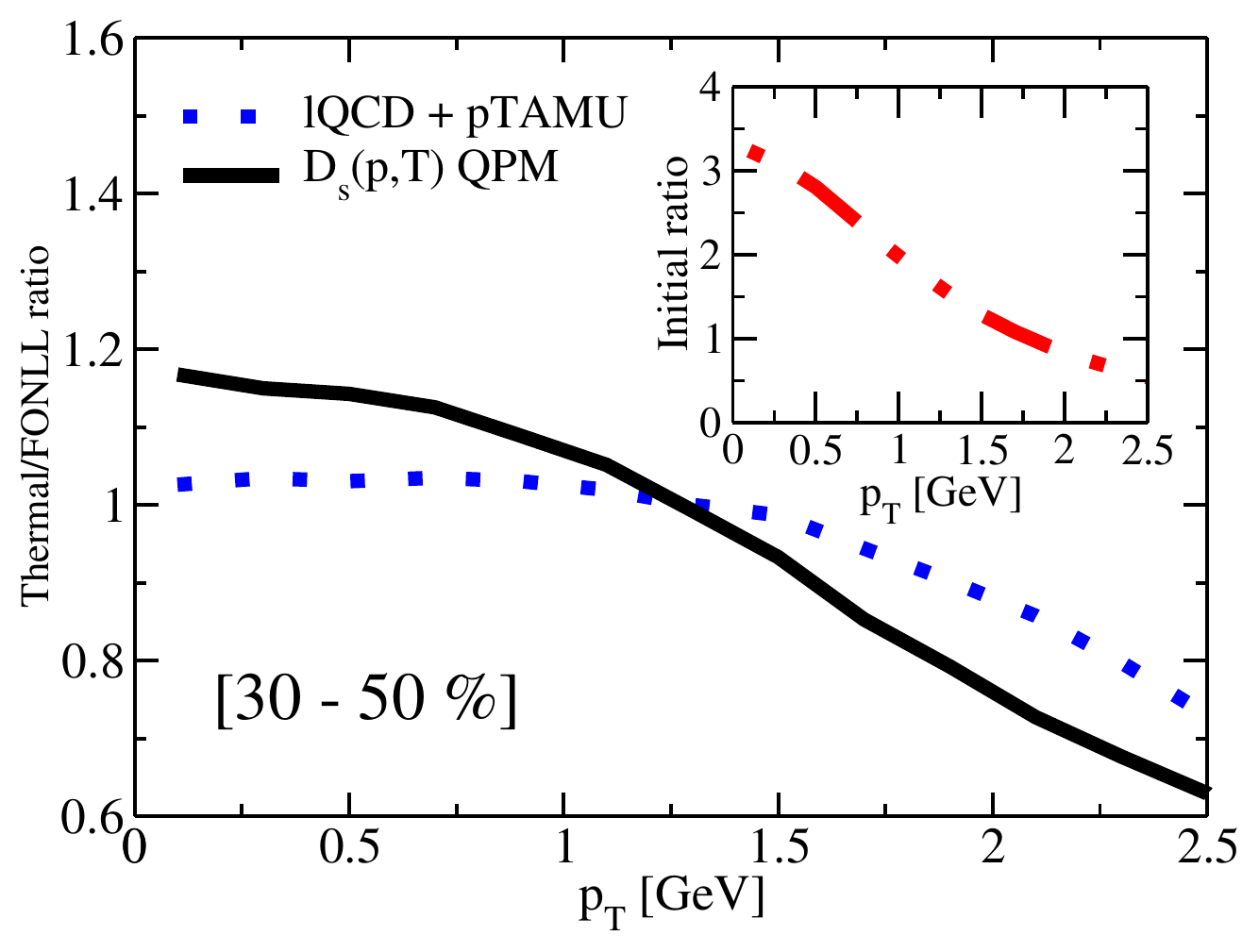}
\caption{Ratio between the final charm quark $p_T$ spectrum obtained from an evolution of charm starting from a FONLL non-equilibrium distribution and the one in which the evolution start from a thermal distribution. The blue dotted line is for the \textit{Case 2} with small $D_s$ and the orange solid line for a larger $D_s$ extracted from standard $QPM$ \cite{Sambataro:2023tlv, Sambataro:2024mkr}. }\label{fig-thermal-ratio}
\end{figure}

The corresponding thermalization time, expressed from the drag coefficient by $\tau_{th}(p)=1/A(p)$, for the different cases studied are shown in fig. \ref{fig-tau}.The cases studied show different momentum dependence: at $p=0$, the \textit{Case 4} (red dot-dashed line) exhibits a thermalization time approximately $50 \, \%$ larger than in the other scenarios but  \textit{Case 1} (solid violet line) of course. In particular, $p_T$ the \textit{Case 4} shows a $\tau_{th}$ which linearly increases with the momentum while the \textit{Case 2} (blue dotted line) has a strong momentum dependence reaching typical lifetime of the fireball already at $p_T \approx 6-8 \, GeV$. This case shows a similar behavior to the \textit{Case 3} (green dashed line) where the momentum dependence comes directly by imposing the FDT, once a constant $D_s$ is assumed. Therefore, it is interesting to note that in the TAMU approach (\textit{Case 2})  which is close to \textit{Case 3} (with $A \sim 1/E$), the $D_s$ is likely to have a very weak p-dependence. In the following, we discuss the impact of the different momentum dependent transport coefficients on the main heavy flavour observables: $R_{AA}$ and $v_{2,3}$. In Fig.s \ref{fig-Raa010} and  \ref{fig-Raa3050}, we show the $R_{AA}(p_T)$ of D mesons at $0\!-\!10 \%$ and $30\!-\!50\%$ centrality classes at midrapidity for $PbPb$ at $\sqrt{s} = 5.02 \, TeV$ in the four cases defined before compared to the available experimental data taken from ALICE collaboration \cite{ALICE:2021rxa} and from CMS collaboration \cite{CMS:2017qjw}. We point out that, the \textit{Case 1} represented by the violet solid line, in which the value of thermalization time is momentum independent and correspond to  $\tau_{th}\approx 1.5 fm/c$, does not give a good description of the experimental $R_{AA}$ inducing a too small $R_{AA}$ approaching to zero in both centrality classes. Notice that this scenario represents a limiting case in which a strong coupling with the medium is imposed across the entire $p_T$ range. With the exception of the scenario just described, all other cases investigated show good agreement with the experimental data also in the high $p_T$ region. Therefore, assuming an interaction strength consistent with the latest lQCD results, only a strong momentum dependence like the one in \textit{Case 2} or \textit{Case 3} can lead to an an $R_{AA}$ that remains comparable with experimental observations for both centralities. Finally, we highlight that the thermalization time $\tau_{th}$ in \textit{Case 4}, corresponding to a weak momentum dependent interaction coming from $QPM_p$ approach with a $D_s$ that at $p=0$ is approximately $50 \, \%$ higher than that extracted from the latest lattice QCD data (Fig. \ref{fig-tau}), also yields a remarkably good agreement with the experimental measurements of $R_{AA}$ in the whole $p_T$ region. The different scenarios studied suggest that the experimental observables are strongly affected by the way in which the interaction between charm and the medium decreases with the increase of momentum of HQ. We can get further constraints from the transport coefficients comparing $v_2(p_T)$ and also $v_3 (p_T)$ for the same cases discussed before with the experimental data \cite{ALICE:2021rxa, CMS:2020bnz}. 
In Fig.s \ref{fig-v2010} and \ref{fig-v23050}, we show the two-particle correlation $v_{2,3}(p_T)$ of D meson for two centrality classes ($0\!-\!10 \, \%$ and $30\!-\!50 \, \%$ respectively) in $PbPb$ at $\sqrt{s} = 5.02 \, TeV$ and midrapidity. The momentum independent \textit{Case 1}, due to the strong interaction with the medium, leads to an enhanced conversion of spatial anisotropy into momentum-space anisotropy, resulting in a $v_2$ strongly exceeding the experimental results. The other three cases lead to a similar $v_{2,3}$, providing a good description of the available experimental data. We obtain a similar trend in both centrality class also for the $v_3$ because it is mainly driven by the initial event-by-event fluctuation, differently by the $v_2$ which is driven by the varying system geometry at different centralities. Note that the differences from \textit{Case 2} to \textit{Case 4} are very small and the uncertainties and discrepancies arising from hadronization and hadronic scattering do not actually allow for a clear distinction among the different cases. Our analysis has shown that assuming a natural p-dependence of the drag, as suggested by FDT and TAMU, the thermalization time of charm quark could be as low as $\tau_{th} \approx 1.5 \, fm/c$ in the low momentum region. 
We conclude this section discussing the implication of the small value of thermalization time predicted by the new lQCD calculation. A value of $\tau_{th} \approx 1 \, fm/c $ at $T_c$ would imply a thermalization time of the charm quark much smaller than the typical lifetime of the fireball. This could entail a nearby complete loss information on the initial stage and initial conditions of $AA$ collisions at low $p_T$. In this direction, we have compared the final charm quark $p_T$ spectra obtained from an evolution of the charm quark starting from a non-equilibrium FONLL distribution to those obtained starting from an initial thermal distribution, by evaluating their ratio. In particular, we evaluate this ratio for the \textit{Case 2} with a small $2\pi T D_s \approx 1$ (at $T_c$) compatible with the new lQCD data and for a larger $2\pi T D_s \approx 4$ (at $T_c$) which comes from standard $QPM$ shown in Fig. \ref{fig-Ds} \cite{Sambataro:2023tlv, Sambataro:2024mkr}. In Fig. \ref{fig-thermal-ratio}, we show the ratio between the charm quark spectra Thermal/FONLL in the two cases. We note that for $p_T < 2 \, GeV$, the resulting ratio turns out to be very close to unity (within a $2 \%$\, up to $p_T \sim M_c$) for the Case 2 with a $2\pi T D_s(p \rightarrow 0) \simeq 1$ implying that in this momentum region charm quarks could have reached a full thermalization or at least observable like final spectra manifest an universal behavior. The ratio in the case of a $D_s$ as in the standard $QPM$ shows a deviation of the order of $20 \, \%$ in the low $p_T$ region.
Hence, the strong interaction suggested by recent lQCD data, paves the way for the idea that even in the heavy-quark sector at low momentum, a dynamical attractor may be reached—leading to a loss of sensitivity of the observables to the system's initial conditions below a certain momentum range.
Therefore, this could imply that charm quarks are expected to show a universal behavior across different systems, where their dynamics can be described by a single scaling parameter, the opacity $\hat \gamma$ \cite{Kurkela:2018qeb,Nugara:2024net}. This means that, despite differences in system size or interaction strength, their evolution follows the same underlying pattern driven by the ratio between the system size and the mean free path. This would also mean that in the low $p_T$ region, charm dynamics could be close to a viscous hydro description, as recently explored in Ref. \cite{Capellino:2022nvf, Capellino:2023cxe}. 

\section{Conclusions}

In this letter, we have studied the propagation of charm quarks in the quark-gluon plasma (QGP) by means of an event-by-event Langevin approach coupled to QGP bulk described by an event-by-event Boltzmann transport approach. 
The study have been developed including the initial state fluctuations by means of a Monte-Carlo Glauber model for the initial distribution of partons to give realistic initial condition of $PbPb$ collisions at $\sqrt{s} = 5.02 \, TeV$. In our simulations, we couple the transport approach with a hybrid hadronization by coalescence plus fragmentation to obtain the final hadron observables. In particular, we have investigated the key observables $R_{AA}$, $v_2$ and $v_3$ of D mesons at top LHC energies, analyzing whether the small values of the new lQCD data for $D_s$ of charm quark can be compatible to the experimental data and focusing on the impact of different heavy-quark interactions with the medium. In this direction, we have analyzed how the description of the above D meson observables is influenced by the momentum dependence of the drag coefficient $A(p)$, which characterizes the strength of HQ-medium interactions as a function of momentum and the related thermalization time $\tau_{th}= A^{-1}$. It is important to notice that the diffusion coefficient $D_s$ extracted provides an estimate of the thermalization time at zero momentum; however, at finite momentum, the interaction strength can signficantly decreases, leading to an increase in $\tau_{th}$. 
Therefore, approaches with different interaction value at $p=0$ can lead to significant different $D_s(T)$ but to comparable $R_{AA}(p_T)$, $v_2(p_T)$ and $v_3(p_T)$ in the whole $p_T$ range below $6 -7 \, \rm GeV$. The results shown in this paper suggest that an extreme scenario with $2\pi T D_s \simeq 1 $ as in lQCD and AdS/CFT, but without momentum dependence, leads to an $R_{AA}(p_T)$, $v_2(p_T)$ and $v_3(p_T)$ out of phenomenology. However, lQCD and AdS/CFT can be compatible with a satisfactory description of $R_{AA}$, $v_{2}$ and $v_{3}$ only assuming a strong momentum dependence of the drag/thermalization time like the one extracted from TAMU approach; quite interestingly this would correspond to a $D_s$ nearly p-independent. Furthermore, a $50 \, \%$ larger $D_s$ but with a weaker momentum dependence like the one in $QMP_p$ scenario, also leads to a very good description of D meson observables within current erros bars and it would imply about $50\%$ larger $\tau_{th}$ at $p=0$ wrt the lQCD data, which is still a quite low thermalization time of about $2  \, fm/c$. Our analysis therefore indicates that, on the $R_{AA}$ and $v_{2,3}$, it is not possible to significantly discriminate between these two cases. A broader investigation involving additional observables or a more advanced statistical analysis would be required, and would be particularly timely and effective
to perform it on the new set of data coming from Run 3 when available also for $v_3$ and for $R_{AA}$ and $v_2$ at different centralities. Finally, we observe that a short thermalization time for charm quarks, $\tau \approx 1 -1.5\, fm/c$ close to $T_c$, as suggested by recent lQCD results, would imply that charm quarks may thermalize within the lifetime of the QGP fireball exhibiting an universal behavior at least up to $p_T \sim M_c$.

\subsection*{Acknowledgments}

We acknowledge the funding from UniCT under PIACERI ‘Linea di intervento 1’ (M@uRHIC) . The authors also acknowledge the support from the European Union’s Horizon 2020 research and innovation program Strong 2020 under grant agreement No 824093 and PRIN2022 (Project code 2022SM5YAS) within Next Generation EU fundings. 



\end{document}